\documentclass[reprint,superscriptaddress,nobibnotes,amsmath,amssymb,aps,prl]{revtex4-2}

\usepackage{graphicx}% Include figure files
\usepackage{dcolumn}% Align table columns on decimal point
\usepackage{bm}% bold math
\usepackage{soul}

\usepackage[colorlinks=true,allcolors=blue]{hyperref}
\usepackage{xcolor}

\newcommand{\refrate}{k_m}

\begin{document}

\title{Mechanical inhibition of dissipation in a thermodynamically consistent active solid}

\author{Luca Cocconi}
\email{luca.cocconi@ds.mpg.de}
\affiliation{%
 Max Planck Institute for Dynamics and Self-Organization, G{\"o}ttingen 37073, Germany
}%
\author{Michalis Chatzittofi}%
\email{mike.chatzittofi@ds.mpg.de}
\affiliation{%
 Max Planck Institute for Dynamics and Self-Organization, G{\"o}ttingen 37073, Germany
}%
\author{Ramin Golestanian}
\email{ramin.golestanian@ds.mpg.de}
\affiliation{%
 Max Planck Institute for Dynamics and Self-Organization, G{\"o}ttingen 37073, Germany
}%
\affiliation{%
 Rudolf Peierls Centre for Theoretical Physics, University of Oxford, Oxford OX1 3PU, UK
}%

\date{\today}

\begin{abstract}
The study of active solids offers a window into the mechanics and thermodynamics of dense living matter. A key aspect of the non-equilibrium dynamics of such active systems is a mechanistic description of how the underlying mechano-chemical couplings arise, which cannot be resolved in models that are phenomenologically constructed. Here, we follow a bottom-up theoretical approach to develop a thermodynamically consistent active solid (TCAS) model, and uncover a non-trivial cross-talk that naturally ensues between mechanical response and dissipation. In particular, we show that dissipation reaches a maximum at finite stresses, while it is inhibited under large stresses, effectively reverting the system to a passive state. Our findings establish a generic mechanism plausibly responsible for the non-monotonic behaviour observed in recent experimental measurements of entropy production rate in an actomyosin material and enzymatic activity in crowded condensates.
\end{abstract}

\maketitle

Active solids constitute a fundamentally out-of-equilibrium class of condensed matter systems, being assembled from units capable of locally converting environmental free energy sources into work. Examples of active solids include crosslinked bio-polymer networks \cite{prost2015active,Koenderink2009,Struebing2020}, confluent cell monolayers \cite{Bi2015,garcia2015physics,ManningPRX2016,alert2020physical}, phase-separated self-propelled colloids with phoretic \cite{palacci2013living,buttinoni2013dynamical} and magnetic \cite{MassanaCid2019} activity, vibrated mechanical networks \cite{shen2016probing}, networks of self-actuated microbots on substrates \cite{baconnier2022selective,Dauchot2024,veenstra2025adaptive}, and hydrodynamically active crystals \cite{Uchida2010a,Oppenheimer2019,tan2022odd}. Models of active solids assembled from active particles, in particular, can be broadly categorized in two classes. The first class corresponds to those systems that undergo an active phase separation arising due to non-equilibrium interfacial (phoretic) interactions \cite{Golestanian2012,AgudoCanalejo2019,Kokkoorakunnel2024} or motility-induced phase separation (MIPS) \cite{Cates2015,Fily2012,Leticia2018} and its variants \cite{mognetti2013living,caprini2023entropons,caprini2023pre}. The second class corresponds to active or motile particles that are coupled via passive elastic linkers and exhibit collective actuation \cite{baconnier2022selective,briand2018spontaneously,deseigne2010collective,xu2023autonomous}, in a manner similar to active polymers \cite{Dreyfus2005,winkler2020physics,bianco2018globulelike,locatelli2021activity,anand2018structure,Kokkoorakunnel2022}.

The frameworks that have so far been developed to study such dense arrays of active particles suffer from a number of key limitations. Firstly, they typically overlook the force-free nature of self-propulsion mechanisms that are designed to work in low Reynolds number conditions \cite{purcell2014life,najafi2004simple,golestanian2005,golestanian2008analytic,lauga2009hydrodynamics}, and, instead, incorporate self-actuation via unphysical ``active force'' monopoles. Secondly, and most importantly for the purpose of this work, they are not thermodynamically consistent, in the sense that they lack a systematic access to the relevant equilibrium limit \footnote{To give but one example, consider the fact that the zero self-propulsion speed limit of hard-core particles with Vicsek-type alignment interactions is in fact \emph{not} an equilibrium one \cite{tasaki2020hohenberg}.}, which can be formally linked to the vanishing of a microscopic affinity and the re-establishment of global detailed-balance. Thirdly, they typically lack a systematic accounting of all relevant contributions to the entropy production, which can for example be dominated by the dissipation in the fluid medium \cite{DaddiMoussaIder2023}, where implementation of momentum conservation will introduce subtleties that affect even the most basic definitions and thermodynamic relations used in probing entropy production \cite{Golestanian2025}. 

A number of works have begun to address these shortcomings by expanding the state space to include one or more explicit chemical cycles \cite{fritz2023thermodynamically,chatzittofi2024entropy,Agranov2024,bebon2024thermodynamics,proesmans2025quantifying,agranov2025entropy}. When these cycles are driven away from equilibrium, currents of the mechanical degrees of freedom emerge, contingent on the existence non-separable terms in the internal energy of the system \cite{Chatzittofi2025a,magnasco1994molecular,challis2013energy,keller2000mechanochemistry,golubeva2012efficiency,jack2020thermodynamic}. 
This non-trivial cross-talk between chemical and mechanical degrees of freedom becomes apparent under the effect of external forces \cite{garcia2017steering}: similarly to how ATPase enzymes (e.g. kinesins) can be made to operate ``in reverse'' as ATP synthases by application of forces exceeding the stall force \cite{liepelt2007kinesin,Golestanian2008a,chatzittofi2024nonlinear}, so can the current associated with the internal chemical cycle of a thermodynamically consistent active particle be inhibited, and even reversed, by the application of a force opposite to the self-propulsion direction \cite{chatzittofi2024entropy}. Perhaps counterintuitively, this dissipative current might also be inhibited by sufficiently large forces parallel to and in the same direction of the swimming direction, a phenomenon known as negative differential mobility \cite{chatzittofi2024entropy,Golestanian2008a,baerts2013frenetic,sarracino2016nonlinear,rizkallah2023absolute}. No such thermodynamically consistent treatment is currently available for active solids. 

These phenomena point to a key feature of active matter thermodynamics, namely, that the rate of free energy dissipation must depend dynamically on the local mechanical environment. In the context of MIPS, for instance, one may wonder to what extent the dissipation of a particle embedded in the condensed phase is inhibited due to bulk stresses. While non-trivial bulk contributions are absent in field theoretic descriptions of MIPS \cite{nardini2017entropy,ro2022model,Markovich2021cost}, formally coarse-grained tight-coupling models point to the existence of a contribution to the local dissipation which is linear in the density \cite{bebon2024thermodynamics}. However, experimental observations suggest that non-linear effects \cite{chatzittofi2024nonlinear} are likely to play a role at intermediate and large densities or stresses \cite{Seara2018Nov, Chen2024Nov}.

\begin{figure}
    \centering
    \includegraphics[width=\columnwidth]{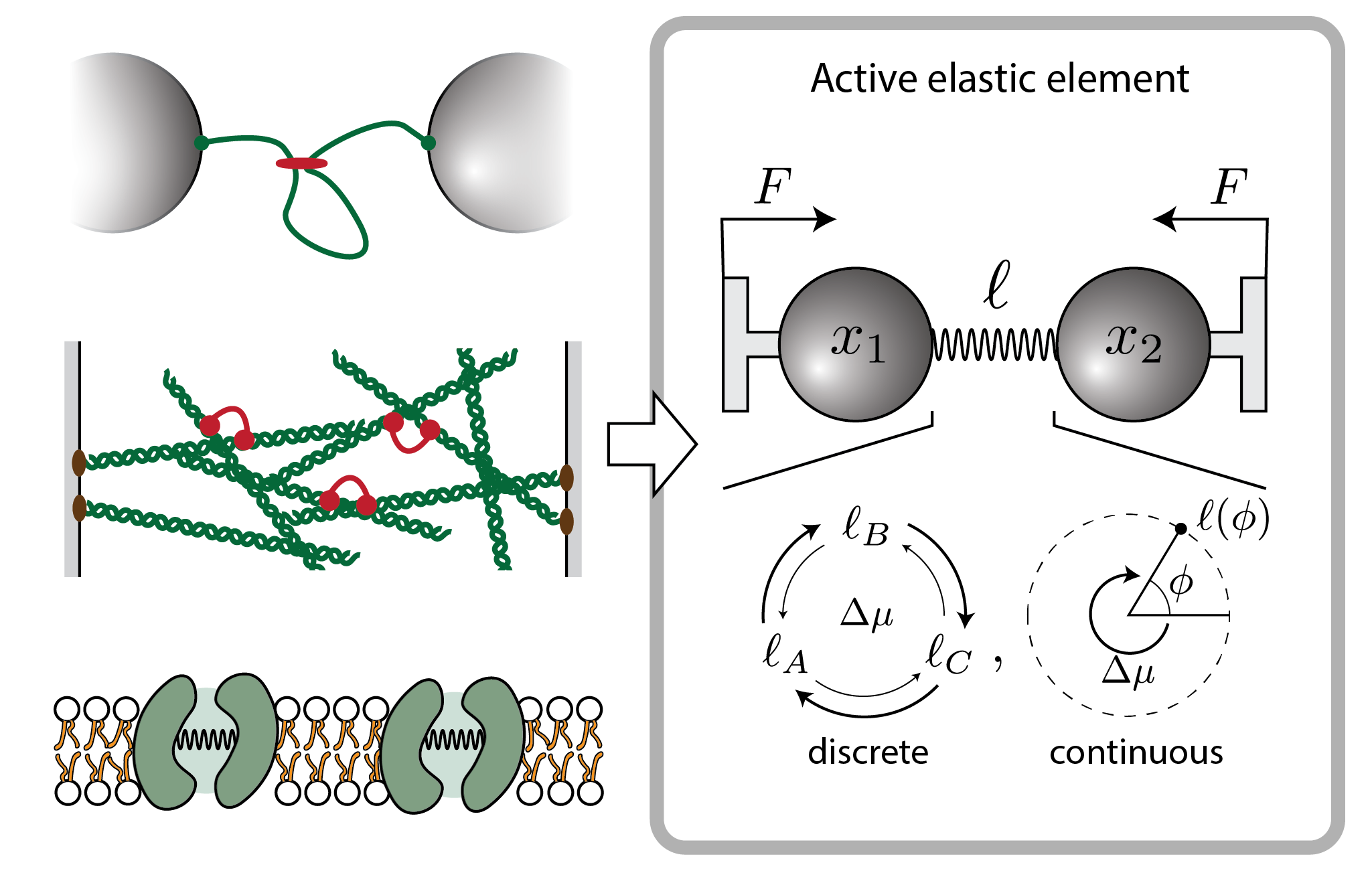}
    \caption{
    The mechanical response to externally applied forces in a broad class of living materials can be represented by the ``active spring'' model introduced in this work. Examples include DNA tethers extruded by ATP-driven cohesin \cite{davidson2019dna}, cross-linked bio-polymer networks (active gels \cite{prost2015active}) and membrane-embedded enzymes \cite{Illien2017Nov,AgudoCanalejo2021,chatzittofi2025,Chatzittofi2025a} or active mechano-sensitive channels \cite{Bonthuis2014}. These applications are further discussed in \cite{suppmat}.
    }
    \label{fig:schematic_bio}
\end{figure}

In this Letter, we establish a thermodynamically-consistent active solid (TCAS) model to explore the non-trivial interplay of mechanics and dissipation in assemblies of strongly-interacting active units. We do so by introducing a new class of active solids, the elementary unit of which is an active elastic element whose dynamics is accounted for in a thermodynamically consistent way. 
This ``active springs" can be effectively realised in a number of biological scenarios, including DNA loop extrusion by cohesin \cite{davidson2019dna}, active gels \cite{prost2015active} and enzyme inclusions \cite{Illien2017Nov,AgudoCanalejo2021,chatzittofi2025,Chatzittofi2025a} or active mechano-sensitive channels \cite{Bonthuis2014} in lipid bilayers, as shown schematically in Fig.~\ref{fig:schematic_bio}. They can also be engineered, e.g. similar to recent experiments on active oscillating nano-motors \cite{Tang2025Jan} or by using DNA-origami \cite{centola2024rhythmically}. Bead-connecting elements undergoing cycles of length variation have been proposed in the context of self-propulsion of synthetic micro-swimmers \cite{golestanian2010synthetic,najafi2004simple,golestanian2008analytic,Golestanian2008a,golestanian2010synthetic}. By considering active elastic elements first in isolation and then assembled into a one-dimensional chain, we show that local dissipation in this system can be inhibited in the presence of sufficiently large contractile and extensile forces. 

\begin{figure*}[t]
    \centering    \includegraphics[width=2.\columnwidth]{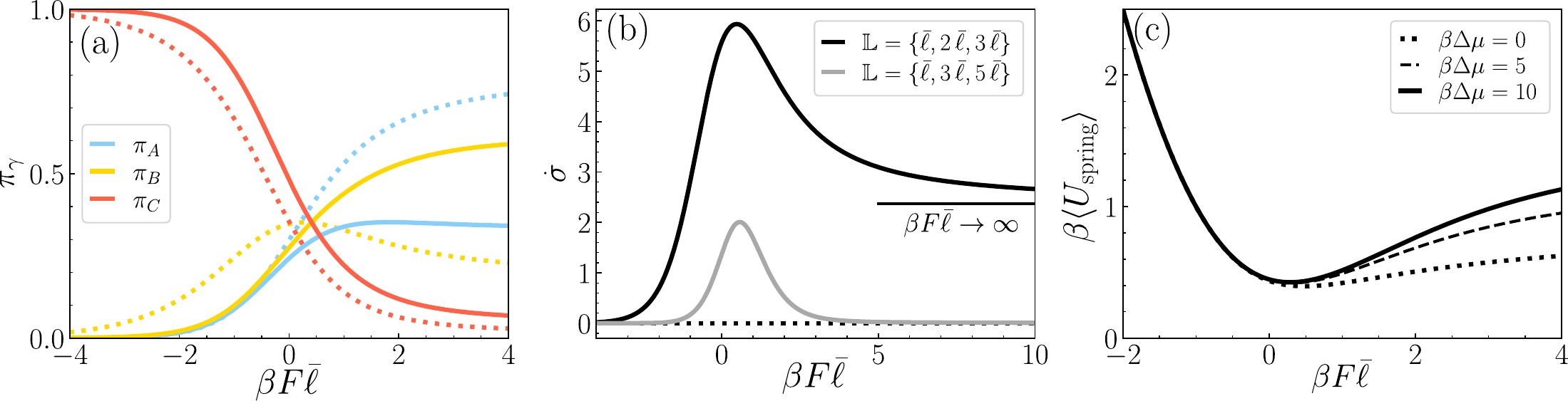}
    \caption{Effect of an external force on a single active elastic element, for $\beta k \bar{\ell}^2=1$ and $U_\ell(\gamma)=0$ for all $\gamma$. In each case, we compare equilibrium ($\Delta\mu=0$, dotted curves) and driven (solid curves) cases: (a) Steady-state probability of the internal state, $\pi_\gamma$, for $\mathbb{L}=\{\bar{\ell},2\bar{\ell},3\bar{\ell}\}$. At equilibrium, the internal state corresponding to the shortest (longest) rest length is generically enhanced under compression (stretching), while driving leads to a repopulation of less energetically favourable states;
    (b) Entropy production $\dot\sigma $ as a function of $F$ for different choice of the accessible rest lengths $\mathbb{L}$. 
    As $F\to \infty$, dissipation reaches a plateau, due to hard core repulsion preventing the inter-particle distance from decreasing below zero;
    (c) Average elastic energy stored in the active elastic element.
    }
    \label{fig:dymer_analytics}
\end{figure*}

\paragraph{The active elastic element---}
Let us consider a dimer of passive Brownian particles connected by an active elastic element in one dimension (Fig.~\ref{fig:schematic_bio}). The positions of the two particles are denoted $x_{1,2}(t)$, while $\ell(t)$ indicates the instantaneous rest length of the elastic element. 
For simplicity, we henceforth assume the associated internal state $\gamma$ to be discrete, with $\gamma \in \{A,B,C\}$ and $\ell_\gamma \in \mathbb{L} = \{\ell_A,\ell_B,\ell_C\}$, noting that three is the minimum number of states required to construct a non-equilibrium chemical cycle \cite{cocconi2020entropy}. Without loss of generality, we introduce a unit of length, $\bar{\ell}$, which represents the typical scale of the rest length of the springs. 
This analysis is generalised to the case of a continuous internal variable in \cite{suppmat}. Upon including the effect of an external (contractile or extensile) force of magnitude $F$, we can write the total energy as
\begin{align}\label{eq:U_def}
 U(x_{1},x_2,\ell) = U_{\rm spring} + U_{\rm rep} + U_\ell + F(x_2-x_1),
\end{align}
where $U_{\rm spring}(x_{1},x_2,\ell) = \frac12 k[(x_2 - x_1) - \ell_\gamma]^2$ is the elastic contribution, $U_{\rm rep}(x_1,x_2)$ arises from short-range hard-core repulsion between the beads, and $U_\ell(\gamma)$ is the energy associated with the internal state of the spring $\gamma$. Denoting $\mu_{\rm p}$ as the single-particle mobility, the Langevin equation for the inter-particle displacement $L \equiv x_2 - x_1$ thus reads
\begin{equation}\label{eq:Ldyn}
    \dot{L}(t) = 2\mu_{\rm p}[-F - k(L - \ell_\gamma) - \partial_L U_{\rm rep}] + \sqrt{4 \mu_{\rm p} k_{\rm B} T} \,\eta,
\end{equation}
where $k_{\rm B} T \equiv \beta^{-1}$ is the thermal energy scale and $\eta$ is a Gaussian white noise of unit covariance, $\langle \eta(t)\eta(t')\rangle = \delta(t-t')$. The rest-length $\ell_\gamma$ evolves according to a Markov jump process with master equation $\dot{P}(\gamma) = M_{\gamma \delta} P(\delta)$, where $M_{\gamma \delta}$ is the transition rate from $\delta$ to $\gamma$. We assume summation over repeated indices. Thermodynamic consistency of the discrete dynamics is ensured by enforcing local detailed balance (LDB) at the level of the transition rate matrix $M$ \cite{maes2021local,chatzittofi2024entropy}.
LDB provides both an interpretation relating probabilistic currents to thermal quantities, such as heat and work flows into reservoirs coupled to the system of interest, and, crucially, a prescription for the dependence of the transition rates on the inter-particle displacement $L$.
Accordingly, we write
\begin{equation} \label{eq:detailedBalance_single}
M_{\gamma \delta}(L) = \frac{\refrate  e^{ \beta \Delta\mu \varepsilon_{\gamma \delta}/6}}{1 + e^{ \beta{\Delta U_{\gamma\delta}(L)}}}, %\pm
\end{equation}
where $\refrate$ is a microscopic rate, whose inverse sets a natural unit of time.  
Here, $\Delta\mu$ is the affinity of the chemical cycle defined to be driven in the clockwise direction $A \to B \to C \to A$, and $\varepsilon_{\gamma \delta}=\pm 1$ depending on the direction of the transition. Moreover, $\Delta U_{\gamma\delta}(L)$ denotes the change in total energy associated with a transition from state $\delta$ to state $\gamma$, i.e. $\Delta U_{\gamma\delta}(L) = \frac{k}{2}\left[ (L-\ell_\gamma)^2 - (L-\ell_\delta)^2 \right] + U_\ell(\gamma) - U_\ell(\delta)$. We thus have $\ln(M_{\gamma \delta}/M_{\delta\gamma}) = \beta [\Delta\mu/3 - \Delta U_{\gamma\delta}(L)]$ and $\ln(M_{\gamma \delta}/M_{\delta\gamma}) = -\beta [\Delta\mu/3 + \Delta U_{\gamma\delta}(L)]$ for clockwise and anti-clockwise transitions, respectively. 

The dynamics given in Eq.~\eqref{eq:Ldyn} satisfies the fluctuation-dissipation relation and, upon setting $\Delta\mu=0$, the joint distribution of $L$ and $\ell_\gamma$ relaxes to an equilibrium steady state with Boltzmann measure $P_{\rm eq}(L,\gamma) \propto \exp[-\beta U(L,\gamma)]$ characterised by zero entropy production. The marginal steady-state probability distribution of the internal state, $\pi_\gamma$, is obtained by integrating $P_{\rm eq}$ with respect to the inter-particle displacement
\begin{align}
    \pi_\gamma &=\frac{1}{Z(F)}\;{\rm Erfc}\left( \sqrt{\frac{\beta}{2  k}} (F-k\ell_\gamma)\right) \nonumber\\ &\times\exp\left(\frac{\beta F}{2k}(F-2k\ell_\gamma) - \beta U_\ell(\gamma) \right), \label{eq:pi_eq_dimer}%\overline \pi(F)\; 
\end{align}
where $Z(F)$ is a force-dependent normalisation factor. Here, hard-core repulsion is taken into account by restricting the support of $P_{\rm eq}$ to $L \in (0,\infty)$. Equation \eqref{eq:pi_eq_dimer} illustrates the importance of thermodynamic consistency, as it leads to a cross-talk between chemical and mechanical degrees of freedom, thereby enabling mechanical forces to  effectively renormalize the energy landscape in chemical space. In particular, in the limit $|F|\to\infty$ we observe an accumulation of probability in the chemical state $\ell_\gamma \in \mathbb{L}$ that maximises $F-2k\ell_\gamma$. For $F<0$, this corresponds to the largest accessible rest length (see Fig.~\ref{fig:dymer_analytics}(a)). Consequently, the probability of states associated with intermediate rest lengths will often be nonmonotonic in $F$. This force-dependent distortion of the chemical energy landscape is further discussed in \cite{suppmat} for the case of continuous $\ell$ (see also \cite{PhysRevLett.87.010602}).

To determine the steady-state joint probability $P(L,\gamma)$ for non-vanishing driving, $\Delta\mu \neq 0$, we perform an adiabatic approximation \cite{pavliotis2008multiscale}. We consider a separation of time scales between fast $L$ and slow $\ell_\gamma$ dynamics, corresponding to the limit $\mu_{\rm p} \to \infty$ in Eq.~\eqref{eq:Ldyn}. The joint probability thus factorises as $P(L,\gamma) = \pi_\gamma P(L|\gamma)$, with $P(L|\gamma)$ being the steady-state conditional probability density of the process \eqref{eq:Ldyn} for constant $\ell(t)=\ell_\gamma$. Here, $P(L|\gamma) \propto {\rm exp}[-\beta(k(L-\ell_\gamma)^2/2 + FL)]$ with support $L \in (0,\infty)$.
After integrating the associated Fokker-Planck equation for $P(L,\gamma)$ with respect to $L$, we obtain an effective Markov jump process governed by the adiabatically-averaged jump matrix $\overline{M}_{\gamma \delta} (F) \equiv \int_0^\infty dL \ M_{\gamma\delta}(L) P(L|\delta)$, which leads to the steady-state solution of the master equation shown in Fig.~\ref{fig:dymer_analytics}(a) for different values of $F$.
With this solution at hand, we can then explore how the dissipation, quantified through the rate of entropy production \cite{seifert2012stochastic,cocconi2020entropy}, as well as the average elastic energy stored in the active spring depend on both $F$ and the driving affinity $\Delta\mu$. In order to calculate the entropy production $\dot{\sigma}(F,\Delta\mu)$ (that we make dimensionless by reporting it in units of $k_{\rm B}$), we draw on Schnakenberg's cycle-based decomposition \cite{schnakenberg1976network}, noting that only the chemical cycle possesses a non-vanishing affinity $\Delta \mu$ \footnote{The possibility to clearly identify the nonequilibrium cycles in a complex states space is another advantage of explicitly enforcing LDB. Phenomenological active models coupling discrete and continuous degrees of freedom may often result in an infinity of cycles of non-vanishing affinity, in which case calculating $\dot{\sigma}$ requires knowledge of the full probability current density.}. 
Thus $\dot{\sigma} = \beta\Delta\mu J_{\rm chem}$ with $J_{\rm chem}(F) = \overline{M}_{B A} \pi_{A} - \overline{M}_{A B} \pi_{B}$ the uniform current in the cycle. The results are shown in Fig.~\ref{fig:dymer_analytics}(b). Remarkably, entropy production displays a non-monotonic dependence on $F$, with a maximum at a small positive force and inhibition for large (positive and negative) forces. Therefore, we find that under sufficiently large forces the active elastic element behaves effectively like a passive spring. 
Additionally, we observe that the average elastic energy $\langle U_{\rm spring}\rangle$ increases in the presence of compressive forces when $\Delta\mu \neq 0$ (see Fig.~\ref{fig:dymer_analytics}(c)), consistent with the intuition that driving allows to populate high energy states that are thermodynamically unfavorable at equilibrium.

\begin{figure}[t]
    \centering
    \includegraphics[width=\columnwidth]{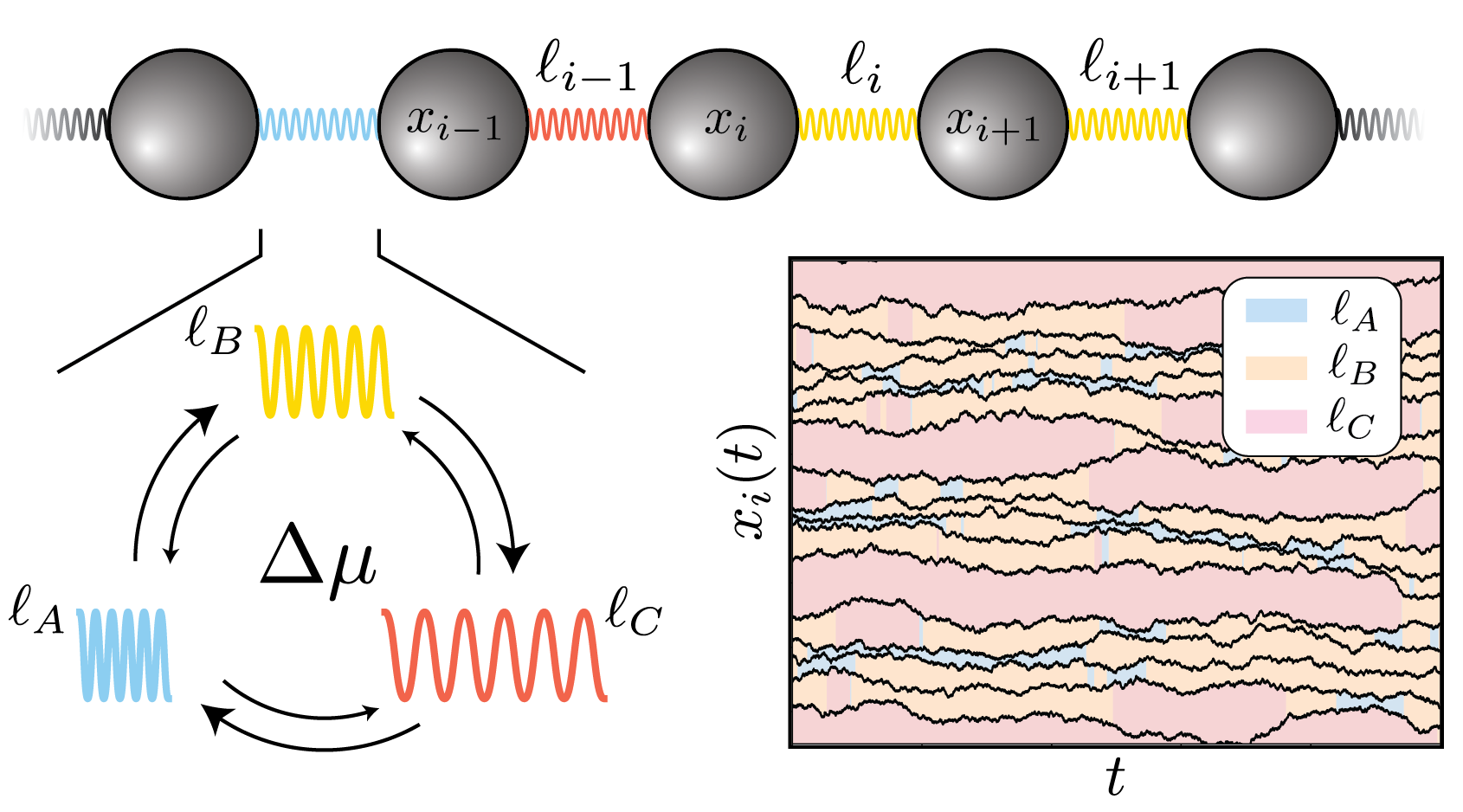}
    \caption{Active elastic elements assembled into a one-dimensional bead-spring chain constitute a TCAS. The inter-particle displacement is controlled by a combination of thermal noise, hard-core repulsion and elastic forces, the latter evolving in time according to the dynamics of the internal chemical state of the springs. A representative realisation of the dynamics is shown, with the area between the trajectories of neighbouring beads (solid black lines) colour-coded according to the instantaneous internal state of the spring.}
    \label{fig:schematic_solid}
\end{figure}

\paragraph{Active solid---}
We now proceed to treat our active elastic element as a fundamental unit from which we assemble a TCAS. We work in one dimension and impose periodic Born-Von Karman boundary conditions, effectively working on a ring of size $R$. A given bead with label $i$, where $i \in \{1, \cdots, N\}$, is located at position $x_i(t) \in [0,R)$ and connected to its left and right nearest neighbours by active springs with internal states $\ell_{i-1}(t)$ and $\ell_i(t)$, respectively, as shown schematically in Fig.~\ref{fig:schematic_solid}. 
Inter-particle displacements are denoted as $L_i = x_{i+1}-x_i$ and satisfy the coupled Langevin dynamics
\begin{align}
\dot{L}_i = 2\mu_{\rm p} K_{ij}(-\partial_{L_{j}}\mathcal{U}) + \sqrt{4\mu_{\rm p} k_{\rm B} T  }\,\eta_i, 
\end{align}
where $\mathcal{U}=\sum_i U^{(i)}(x_i,x_{i+1},\ell_{\gamma_i})$ and $ \eta_i(t)$ is a Gaussian white noise that is nearest-neighbour correlated, i.e. $\langle \eta_i(t) \eta_j(t') \rangle = K_{ij} \delta(t-t')$, with $K_{ij}\equiv \delta_{ij}-\delta_{\langle ij\rangle}/2$.  
The nondiagonal structure of the mobility matrix $K$ is essential to ensure $\sum_i L_i=R$ under the stochastic dynamics.
The contribution to the total energy from each spring, $U^{(i)}$, is defined analogously to Eq.~\eqref{eq:U_def}, although in this case we assume that no external forces act on the chain.
All chemical cycles are driven by an identical affinity $\Delta\mu$. The transition rate matrix governing the dynamics of the $i$th spring, $M^{(i)}_{\gamma_i\delta_i}$, independently satisfies LDB along the lines of Eq.~\eqref{eq:detailedBalance_single}, 
with $\Delta U_{\gamma \delta}$ replaced by $\Delta U^{(i)}_{\gamma_i \delta_i}$. See Appendix A %\ref{sec:coarse_grainining_chain} 
for explicit expressions.
Finally, hard core repulsion between nearest neighbour beads is described via a Weeks-Chandler-Anderson potential \cite{weeks1971role}. We will now examine the behaviour of this spatially extended model using numerical simulations (details of the simulation algorithm are given in \cite{suppmat,moller2003statistical}).

\begin{figure}[h!]
    \centering
    \includegraphics[width=1.07\columnwidth]{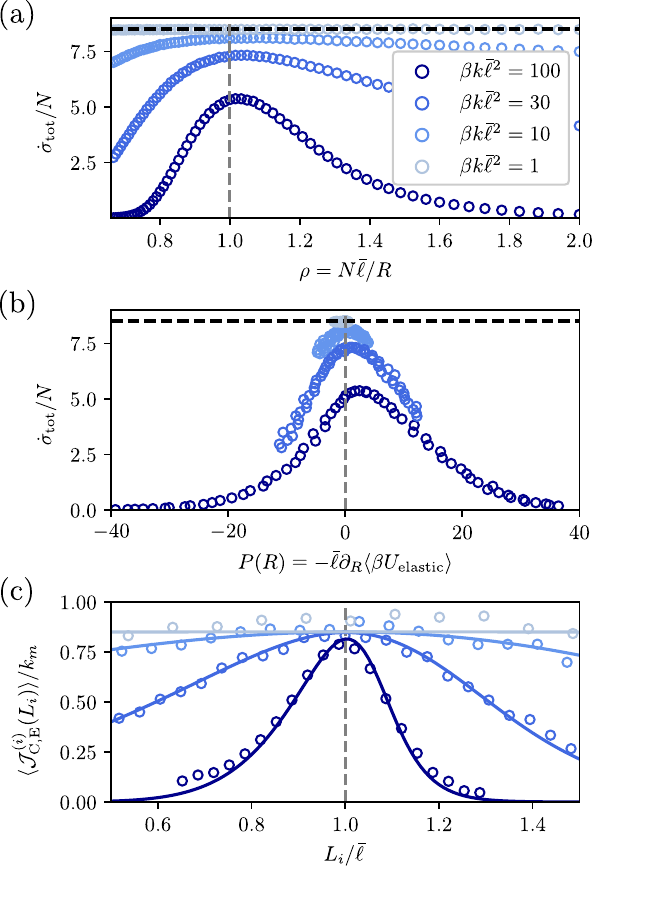}
    \caption{Mechanical inhibition of dissipation in a TCAS: (a) average entropy production per bead as a function of the dimensionless bead density $\rho = N\bar{\ell}/R$, controlled by varying the ring size $R$ at fixed $N$, for different values of the elastic constant $k$. Here, we take $\beta \Delta\mu=10$, $\mu_{\rm p}=0.01\beta\refrate\bar{\ell}^2$, $\beta \epsilon=1$, $\ell_A = 0.9\,\bar{\ell}$, $\ell_B=\bar{\ell}$ and $\ell_C=1.1\,\bar{\ell}$. The length scale associated with the WCA potential is set to $a=0.1\,\bar{\ell}$.
    The horizontal dashed line corresponds to $\dot{\sigma}_{\rm tot}/N = \frac13 \beta\Delta\mu \sinh(\beta\Delta\mu/6)$, as obtained from Eq.~\eqref{eq:detailedBalance_single}
    with $\Delta U_{\gamma\delta} =0$ for all pairs $\gamma,\delta \in \mathbb{L}$; 
    (b) average entropy production as a function of the effective dimensionless pressure \cite{suppmat}; 
    (c) average conditional current in the dissipative cycle of a given active spring (see \cite{suppmat} for definitions) as a function of the associated spring length $L_i$ for $N=64$ and $R=64\,\bar{\ell}$ [all other parameters as in (a)]. Solid curves indicate the analytical expression \cite{suppmat}. }
    \label{fig:epr_uelastic_chain}
\end{figure}

The total average rate of entropy production is found as $\dot{\sigma}_{\mathrm{tot}} = N J_{\mathrm{chem}} \beta \Delta\mu$ where $J_{\mathrm{chem}}$ is the steady-state probability current of any given chemical cycle \cite{zeeman1988stability,huang2015steady}. 
This quantity is plotted in Fig.~\ref{fig:epr_uelastic_chain}(a) as a function of the dimensionless bead number density $\rho \equiv N \bar{\ell}/R$, which we control by varying the periodic domain size $R$. 
Crucially, we observe a non-monotonic dependence of $\dot{\sigma}$ on the bead density, with a maximum at a characteristic spacing $\rho^{-1} = {\rho^*}^{-1}$ commensurate with the set of accessible values of the rest length ($\mathbb{L}$). For $\rho \ll \rho^*$, i.e. upon stretching, $\dot{\sigma}$ decays exponentially to zero, while a slower dropoff is observed for $\rho \gg \rho^*$, i.e. upon compression (see also Fig.~S4 in \cite{suppmat}). 
Clearly, this \emph{mechanical inhibition of dissipation} originates from the cross-talk between mechanical and chemical degrees of freedom induced, at the level of each spring, by the presence of a non-separable contribution to the internal energy in the form of $U^{(i)}_{\rm elastic} = \frac12 k(L_i -\ell_{\gamma_i})^2$. Accordingly, we can weaken this effect by reducing the elastic constant $k$, with mechanochemical coupling being fully eliminated upon setting $k=0$. In this limit, the linear scaling of the total dissipation with density is recovered \cite{bebon2024thermodynamics}.
Analogous results are obtained by plotting $\dot{\sigma}/N$ against the effective pressure; see Fig.\ref{fig:epr_uelastic_chain}(b). The effect of activity on the associated compressibility is discussed in \cite{suppmat}.
To establish the locality of this inhibitory effect, we additionally calculate the average chemical current associated with any given elastic element \emph{conditioned} on a specific value $L_i$ of its length (Fig.\ref{fig:epr_uelastic_chain}(c); see \cite{suppmat} for definitions). In line with the result for the total entropy production, we identify a peak at $L_i^* \simeq \ell \in \mathbb{L}$ which quickly broadens with decreasing $k$, indicating that the effective chemical mobility becomes less sensitive to the spring length.
Increasing (decreasing) the typical values of $\ell$ in $\mathbb{L}$ shifts the dissipation peak towards larger (smaller) values of $L_i$ (see \cite{suppmat}).
The $\rho$-dependence of the elastic energy stored in the active solids and the probability of observing a particular value of $\ell_{\gamma_i} \in \mathbb{L}$ are discussed in \cite{suppmat}. Comparison of Figs.~\ref{fig:dymer_analytics} and \ref{fig:epr_uelastic_chain} suggests an analogy between the bead number density of the periodic chain and the external force applied to an isolated dimer. This correspondence can be made exact by performing an explicit coarse-graining, as shown in Appendix A.

A non-monotonic dependence of the average entropy production on the persistence time (non-dimensionalised by the stiffness $k$) was previously reported in Ref.~\cite{caprini2023entropons} for a non-thermodynamically-consistent active bead-spring model. We note, however, that the aforementioned observation does not stem from an underlying mechanochemical coupling; it originates from the combined effect of confinement and inertia \cite{frydel2023entropy}. 

\paragraph{Conclusion---} 
We have introduced a class of TCASs assembled from elementary active elastic elements, whose rest-length dynamics are controlled by a dissipative chemical cycle. This setup allows us to illustrate a fundamental property of strongly interactive active matter, namely, the non-trivial interplay between their mechanics and thermodynamics. In particular, starting from generic (yet often misrepresented) ingredients, we demonstrate that dissipation can be inhibited in the presence of large stresses, effectively reverting the active material to a passive state. The entropy production rate exhibits a non-monotonic behaviour as a function of the stresses, and in particular, reaches a maximum at some non-vanishing intermediate force strength. 
Our framework is in good agreement with direct experimental measurements of the energy dissipation in crowded systems of myosins that generate active stresses, serving as a minimal model for similar systems \cite{Seara2018Nov}. It also introduces a potential mechanism behind recent observations of non-monotonic enzymatic activity in crowded condensates \cite{Dindo2025Jan}.
Our predictions can be further tested in a number of biological systems, including DNA tethers, bio-polymer networks, and biological membranes (see Fig.~S\ref{fig:schematic_bio} in \cite{suppmat}).
In this respect, we note that, while direct measurements of entropy production are generally challenging \cite{chatzittofi2024entropy,diterlizzi2024variance}, in cases where dissipation stems from nonequilibrium catalysis of a chemical reaction of known affinity, the yield of one or more product species can be used as a proxy to estimate the chemical current and thus dissipation in the cycle \cite{Dindo2025Jan}.

While hydrodynamic effects were ignored in the discussion above, we have verified that our findings are not affected qualitatively for sufficiently small systems when they are included, as detailed in \cite{suppmat}. 

Due to the importance of out-of-equilibrium phase separation in many biological functions \cite{brangwynne2009germline,sokolova2013enhanced,hyman2014liquid,zwicker2017growth}, characterising the role of mechanical inhibition of activity in the bulk of active matter condensates remains a fundamentally important open question. From an engineering perspective, a more rigorous understanding of mechanochemical coupling in active solids could also be leveraged to design ``smart'' mechanosensitive materials capable of generating chemical signals in response to local mechanical perturbations \cite{haswell2011mechanosensitive,garcia2017steering}. In particular, the ability to control the location of the dissipation peak by engineering a suitable $\mathbb{L}$ and $M$ could be exploited to design materials that catalyze specific sets of reactions only under defined mechanical conditions.

In addition to providing a template for further studies of the thermodynamics of active solids in higher dimensions, the 1D passive-active bead-spring model studied here can be considered as a prototype for a thermodynamically consistent active polymer once embedded in 3D space (cf.~\cite{winkler2020physics,bianco2018globulelike,locatelli2021activity,anand2018structure}). 

\begin{acknowledgements}
LC and MC thank Connor Roberts and Daniel Seara for useful discussions. We acknowledge support from the Alexander von Humboldt Foundation and from the Max Planck School Matter to Life and the MaxSynBio Consortium, which are jointly funded by the Federal Ministry of Education and Research (BMBF) of Germany and the Max Planck Society.
\end{acknowledgements}

\bibliography{updated_bibliography,Golestanian}

\section*{End Matter}

\paragraph{Appendix A: Coarse-graining of the chain---}%
In order to map the dynamics of the active solid to a single active dimer, we define the probability distribution $\mathcal{P}_N(L_{1},...L_{N},\gamma_1,...\gamma_N,t)$,  
which evolves according to the Fokker-Planck equation
\begin{equation*}
    \partial_t \mathcal{P}_N = \sum_i^N \left[ \sum_{\delta_i } M^{(i)}_{\gamma_i \delta_i} \mathcal{P}_N (...,\delta_i,...) -  \partial_{L_i}J_i \right],
\end{equation*}
where 
\begin{align*}
     \mu_{\rm p}^{-1}J_i &= -\beta^{-1}\partial_{L_i}\mathcal{P}_N - k [\Delta L_{i+1} 
    + \Delta L_{i-1} 
    - 2\Delta L_{i} ]\mathcal{P}_N\\ &- [\partial_{L_{i+1}} U_{\rm rep}^{(i+1)}+ \partial_{L_{i-1}} U_{\rm rep}^{(i-1)} - 2 \partial_{L_{i}} U_{\rm rep}^{(i)} ]\mathcal{P}_N,
\end{align*}
and $\Delta L_{i} \equiv L_i - \ell_{\gamma_i}$. The transitions rate matrices are
\begin{equation} %{\refrate e^{\pm \beta \Delta\mu/6}}
    M^{(i)}_{\gamma_i \delta_i}(L_i) = \frac{\refrate e^{ \beta \Delta\mu \varepsilon_{{\gamma_i} {\delta_i} }/6}}{1 + e^{ \beta{\Delta U^{(i)}_{{\gamma_i} {\delta_i} }}}} ,
\end{equation}
where
\begin{equation}
    \Delta U^{(i)}_{\gamma_i\delta_i}(L_i) = \frac{k}{2}\left[ (L_i-\ell_{\gamma_i})^2 - (L_i-\ell_{\delta_i})^2 \right] + U_\ell(\gamma_i) - U_\ell(\delta_i).
\end{equation}
The total state space is thus $\mathbb{S} = [0,R)^{\otimes N} \otimes \mathbb{L}^{\otimes N}$. 
We now integrate out $2(N-1)$ coordinates associated with all but one elastic element as defined below
\begin{align}
    \mathcal{P}_1(L_i,\gamma_i) = \left( \prod_{j \neq i}^N \sum_{\gamma_j} \int dL_j \right) \mathcal{P}_N,
\end{align}
and truncate the resulting BBGKY hierarchy by using the closure approximation $\mathcal{P}_2(L_1,L_2,\gamma_1,\gamma_2) \simeq \mathcal{P}_1(L_1,\gamma_1)\mathcal{P}_1(L_2,\gamma_2)$. We obtain
\begin{align}\label{eq:reduced_fp_cg}
    \partial_t \mathcal{P}_1 &= \sum_{\delta=1}^3 M_{\gamma \delta}\mathcal{P}_1 + \beta^{-1} \mu_{\rm p}\partial_L^2 \mathcal{P}_1  \nonumber \\
    &+ 2 \mu_{\rm p}\partial_L \big([F_1-k(L-\ell_\gamma) -\partial_L U_{\rm rep} ]\mathcal{P}_1\big),
\end{align}
where
\begin{align}\label{eq:f_reduced}
    F_1 = \sum_{\delta}\int dL' [-k(L' - \ell_\delta)-\partial_{L'}U_{\rm rep}] \mathcal{P}_1(L',\delta).
\end{align}
We thus observe that Eq.~\eqref{eq:reduced_fp_cg} is takes on the form of an effective single-dimer Fokker-Planck equation (presented in Eq.~\eqref{eq:Ldyn}) with an external force $F_1$. Equations \eqref{eq:reduced_fp_cg} and \eqref{eq:f_reduced} can be solved self-consistently to obtain the exact relationship between $F_1$ and the bead number density. It is intuitively clear that large densities will result in strong compressive forces, and vice versa.

\end{document}